\documentclass[aps,prl,twocolumn]{revtex4}
\usepackage{amsmath}
\usepackage{amssymb}
\usepackage{graphics}
\usepackage{epsfig}
\usepackage{color}
\usepackage{graphicx}
\usepackage{color}
\begin{document}
\title{Rotation induced grain growth and stagnation in phase field crystal models}
\author{Mathias Bjerre$^1$, Jens Tarp$^1$, Luiza Angheluta$^2$, Joachim Mathiesen$^1$}

\affiliation{$^1$ Niels Bohr Institute, University of Copenhagen, Blegdamsvej 17, DK-2100 Copenhagen, Denmark. \\
$^ 2$ Physics of Geological Processes, Department of Physics, University of Oslo, Norway}

\begin{abstract}
We consider the grain growth and stagnation in polycrystalline microstructures.  From the phase field crystal modelling of the coarsening dynamics, we identify a transition from a grain-growth stagnation upon deep quenching below the melting temperature $T_m$ to a continuous coarsening at shallower quenching near $T_m$. The grain evolution is mediated by local grain rotations. In the deep quenching regime, the grain assembly typically reaches a metastable state where the kinetic barrier for recrystallization across boundaries is too large and grain rotation with subsequent coalescence is infeasible. For quenching near $T_m$, we find that the grain growth depends on the average rate of grain rotation, and follows a power law behavior with time with a scaling exponent that depends on the quenching depth.
\end{abstract}
\maketitle

\emph{Introduction} -- Polycrystalline micro-structures are typically formed by thermal processes such as quenching or annealing of melts, through the nucleation and growth of grains of different crystallographic orientation. Since these micro-structures have a controlling role on the large scale material properties, e.g.  mechanical, magnetic and optical properties, yield strength, it is crucial to understand their formation and late-stage evolution. 

Curvature driven grain growth is described  in two dimensions (2D) by the von Neumann-Mullins growth law and  predicts a linear growth of the average grain area, $\langle A\rangle \sim t$, that follows directly from the linear relationship between grain boundary normal velocity and curvature~\cite{neumann1952, mullins1956two}. However, one common intriguing findings in experiments with thin metallic films and molecular dynamics simulations of annealed polycrystalline systems is that the coarsening deviates substantially from the curvature-driven growth~\cite{humphreys2004recrystallization,holm2010grain}. Instead, the grain area size increases as $\langle A \rangle \sim t^{\alpha}$, where the value of the scaling exponent $\alpha$ may depend non-trivially on a number of controlling factors, such as the annealing temperature, grain size, grain boundary mobility and surface energy. For instance, the growth kinetics in nanocrystalline Fe was experimentally observed to be controlled by the grain size leading to a super-diffusive growth, $\alpha \approx 2$~\cite{krill2001size}. However, at mesoscales, the coarsening law in metallic thin films is typically sub-diffusive $\alpha\approx 1/2$~\cite{nichols1993situ,barmak2013grain}. A unified theoretical foundation for the anomalous grain growth is still lacking, despite numerous attempts to generalize the phenomenological normal growth model to include additional mechanisms that control the growth rate~\cite{cahn2004unified,moldovan2002scaling}. 

Grain growth stagnation is another anomalous behavior  to the classical picture of a curvature-driven scaling law and occurs in a wide range of materials ranging from metallic thin films~\citep{frost1994microstructural,barmak2013grain} to ice~\citep{de1987grain}. Although, the thermodynamically stable state consists of a single crystal, the kinetics towards equilibrium goes through an energy landscape with possible metastable states where grain growth stagnates. Several mechanisms have been identified to cause stagnation such as boundary pinning by impurities, the presence of a boundary melt or thin films between grains~\citep{holm2010grain}. Experiments on soap froth have revealed an intermittent grain boundary dynamics that is hard to reconcile with the classical picture of a uniform grain growth~\citep{macpherson2007neumann,schmidt2004watching}. Based on molecular dynamics simulations~\cite{holm2010grain}, it has been argued that the grain-boundary roughness controls the boundary mobility hence the overall grain growth, and that the presence of a small fraction of low-mobility, smooth grain boundaries can lead to stagnation. Grain rotation and grain coalescence due to a coupling between the normal growth and tangential motion have recently been suggested as an important processes in the evolution of high purity polycrystalline materials~\citep{cahn2004unified,moldovan2002scaling}. The experimental observations on grain rotation, have further been backed by models reproducing the sub-diffusive scaling behavior of grain growth dominated by grain rotation and coalescence~\cite{moldovan2002scaling}.

\begin{figure}
\includegraphics[width=.45\textwidth]{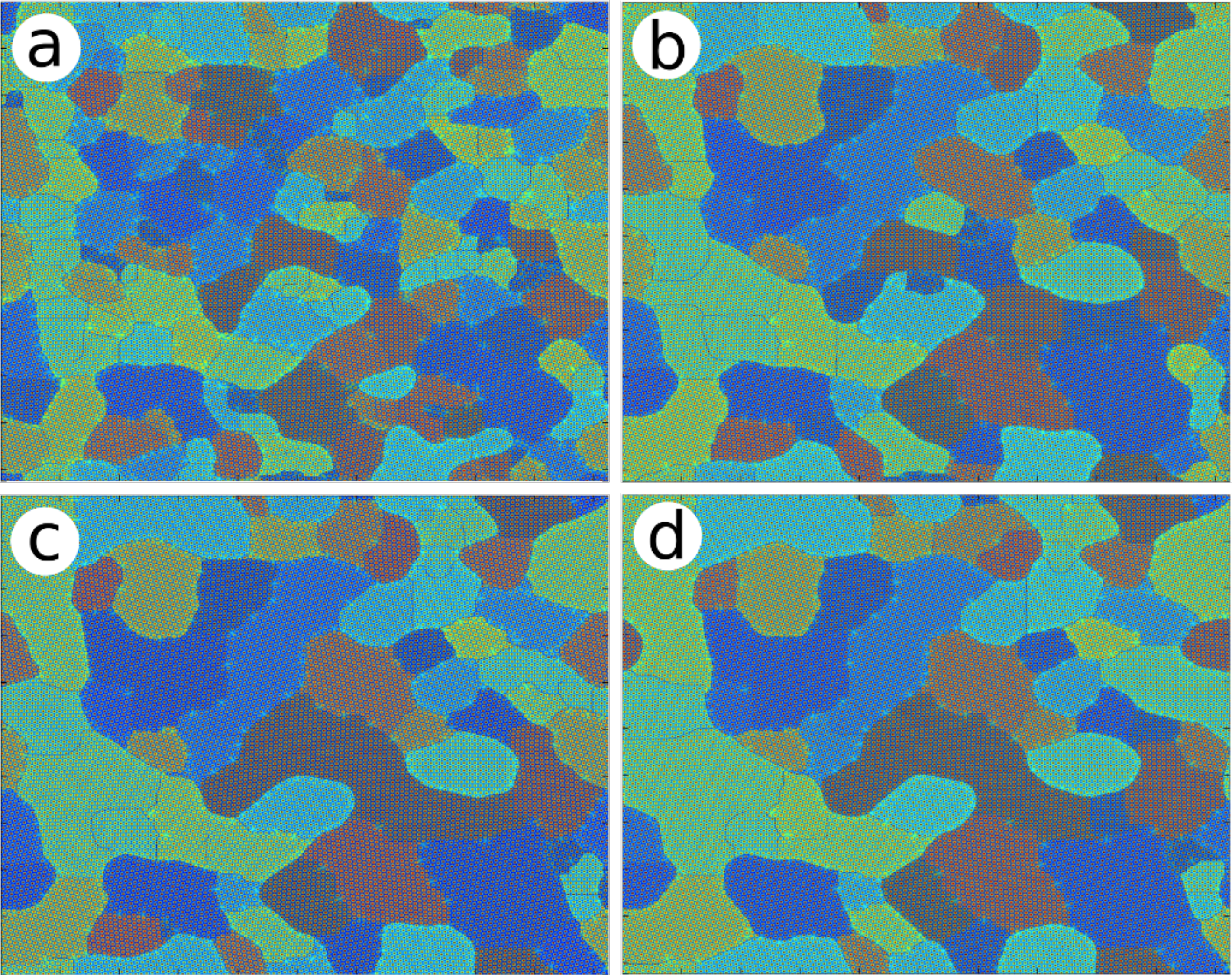}%
\caption{Time snapshots of grain growth in a polycrystalline material taken at simulation timesteps (a) $t=400$, (b) $t=12000$, (c) $t=36000$ and (d) $t=225000$.  The system size is 1024 by 1024 which corresponds to approximately 100 by 100 atom units and the simulation has been run with the temperature controlling parameter set to $a_2=-0.05$  (corresponding to a relatively cold system). The individual grain colors denote the lattice orientation of the individual crystallites. In the last frame the grain growth has reached a fixed state where even relatively high levels of thermal noise cannot reactivate grain boundary migration or internal rearrangement.}\label{timevolution}
\end{figure}

\begin{figure}
\includegraphics[width=.48\textwidth,clip]{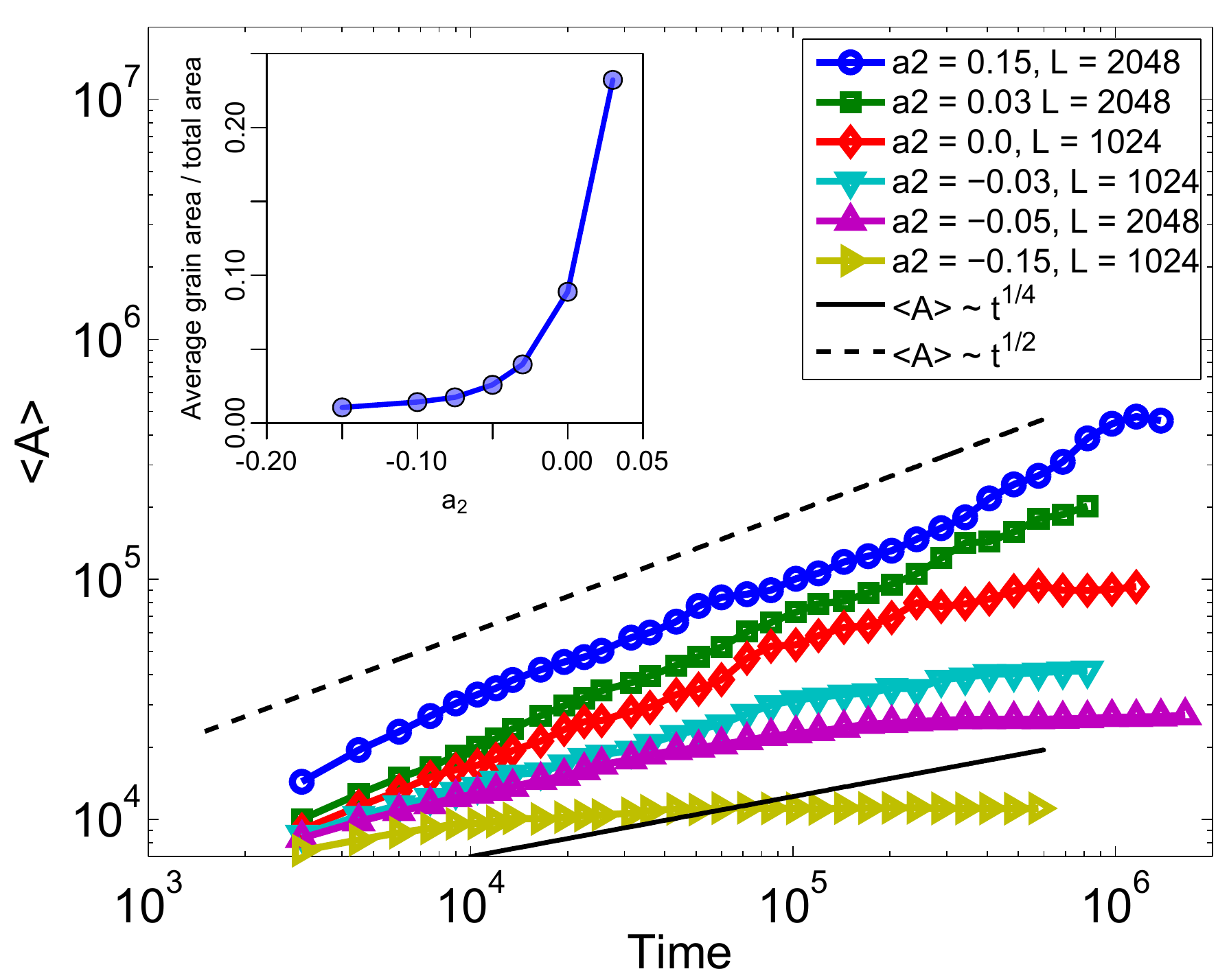}
\caption{Mean grain size as a function of time. We observe a characteristic power law scaling with a temperature dependent scaling exponent for the mean grain size as function of time. For cold temperatures $a_2\approx -0.1$ the scaling exponent is close to $\alpha =1/4$ whereas for higher temperatures it continuously changes towards a value of $\alpha=1/2$. The inset shows the average grain area in the stagnated state for various temperatures (values of $a_2$) and for a system of size $L=2048$.}\label{scaling}
\end{figure}

In this Letter, we present numerical results on the anomalous coarsening dynamics of polycrystalline films and the influence of grain rotation. We use the phase field crystal model that has been shown to be an efficient approach to modelling various aspects of polycrystalline dynamics on diffusive timescales~\cite{elder2007phase,mellenthin2008phase,stefanovic2006phase,adland2013unified}. 
We observe a transition between a dynamical state where grains continuously coarsen and a state where grain growth stagnates, with a cross-over time that diverges as the quenching depth is lowered. Our setup consists of initial crystal seeds of random lattice orientations and uniformly scattered in a two-dimensional undercooled melt. During crystal growth, we track the lattice orientation and the area occupied by the individual grains. Fig.~\ref{timevolution} shows a few snapshots of polycrystalline textures during a coarsening process. At early stages in the coarsening process, the mean grain area scales with time with a power-law exponent that depends on quenching temperature and system size. This non-universal scaling regime may survive on longer timescales only for shallow undercoolings below the melting temperature $T_m$, whereas it crosses over to a stagnation plateau when the melt is deeply quenched to temperatures much lower than $T_m$. The saturation value of the mean grain area depends non-trivially on the quenching temperature and system size (see Fig.~\ref{scaling}).

\emph{The model} -- The phase field crystal (PFC) model operates on microscopic lengthscales and diffusive timescales, and thus constitutes an efficient computational alternative to traditional atomistic methods that are constrained on short timescales comparable to atomic vibrations. The PFC method has been applied to different nonequilibrium phenomena in (poly)crystalline materials including phase transitions~\cite{elder2007phase,Achim06}, elastic and plastic deformations~\cite{stefanovic2006phase,emmerich2012phase}. The coarse-grained time resolution of the PFC method allows for an efficient modelling of slow dynamics of dissipative structures such as grain boundaries and crystal defects. In the simplest formulation, the evolution of the PFC density field $\psi$ is governed by an over-damped, diffusive  equation of motion on the form
\begin{equation}\label{eq:PFC}
\frac{\partial\psi}{\partial t}=\nabla^2\frac{\delta\mathcal{F}[\psi; T]}{\delta\psi}
\end{equation}
where the static free energy functional $\mathcal{F}[\psi; T]$, that determines the equilibrium properties of the crystal phase, has the  phenomenological form of the Swift-Hohenberg free energy, but can also be derived from microscopic details using the density functional theory~\cite{elder2007phase}. Here, we consider the free energy functional as derived from the density functional theory for a hexagonal (fcc) crystal lattice in 2D and given as
\begin{eqnarray}\label{eq:PFC_F}
&&\mathcal{F}[\psi ; T] =\nonumber\\
&& \int d\mathbf{r} \left[\frac{1}{2} \psi \left( \nabla^{2}+1 \right)^{2}\psi + \frac{a_2}{2}\psi^{2} -\frac{1}{6}\psi^{3} + \frac{1}{12} \psi^{4} \right],
\end{eqnarray}
where $a_2$ is proportional to the quenching depth relative to the critical melting temperature $T_m$. The local terms correspond to the coarse-grained free energy of an ideal gas, whereas the non-local term follows from the lowest-order gradient expansion of the interactions that allow for a periodic ground state corresponding to a triangular lattice in 2D. 

\begin{figure}
\includegraphics[width=.48\textwidth]{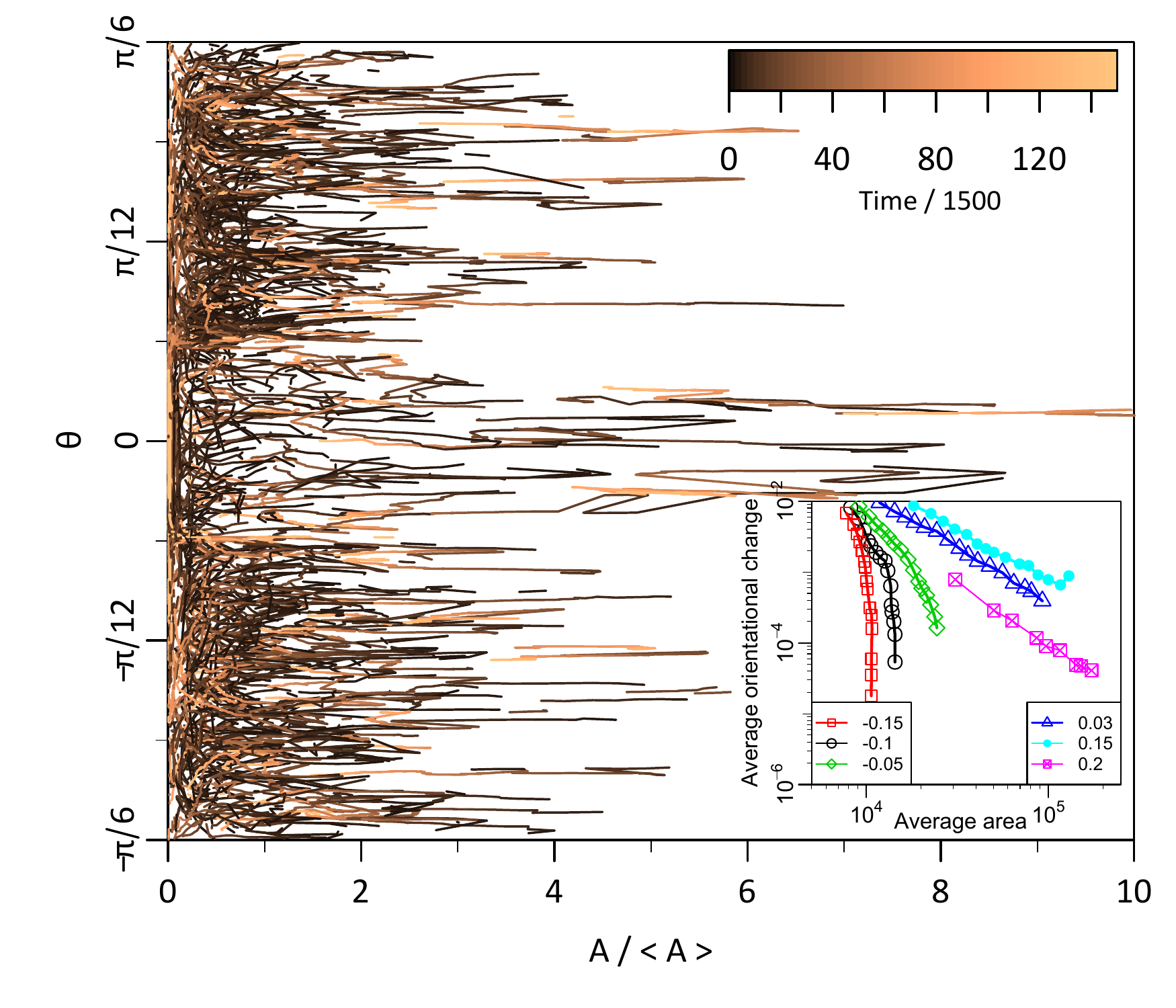}
\caption{(Color online) Individual grain trajectories in a lattice orientation and grain area diagram. The areas of the individual grains have been normalized by the average size to a given time. The time evolution along the trajectories is indicated by the color legend. Inset, the average change in grain orientation versus the average grain area. At low temperatures (low values of $a_2$), grain rotation and grain growth quickly stagnates. Close to the melting temperature  grain rotation goes down while the grain growth remains fast.}\label{graindist}
\end{figure}

Several extensions of Eq.~(\ref{eq:PFC}) have been proposed by introducing additional timescales for faster acoustic relaxation of the elastic fields. The modified PFC model proposed in Ref.~\cite{stefanovic2006phase} is based on a two-time-scale dynamics of the PFC mimicked by a second order time-derivative of the $\psi$-field, where the fast timescale resolves the rapid elastic relaxation. The fast dynamics related to phonons is taken into account by a three-timescale dynamics of the PFC that can be derived from a generalized hydrodynamics of solids~\cite{majaniemi2007}. Since polycrystalline dynamics is a dissipative process, the fast elastic relaxation timescale does not influence the coarsening dynamics. Thus, for simplicity we present numerical results obtained from the diffusive PFC given by Eq.~(\ref{eq:PFC}), although similar results where achieved from test simulations including elastic relaxations.  

We solve numerically Eq.~(\ref{eq:PFC}) using a pseudo-spectral method similar to that in Ref.~\cite{mellenthin2008phase}. All simulations were carried out with an average density equal to that of the solid phase predicted by the one-mode approximation and using the common tangent construction. Simulations were carried out at different quenching depths and crystal densities. The parameter $a_2$ was chosen in the range $-0.15$ to $0.2$ which is below the liquid transition which appears at $a_2=0.25$. 

The individual crystal orientations can be extracted from the phase field crystal density by using a wavelet transformation~\cite{singer2006analysis}. Calculating the magnitude of the gradient in the grain orientation, we observe that the orientation changes most rapidly across grain boundaries. We can use this together with a watershed algorithm to identify individual grains, (see Fig.~1). 

\emph{Results and discussion} --Polycrystalline microstructures, formed by grain nucleation and growth from an undercooled melt, initially coarsen with time according to a power law 
\begin{equation}
\langle A \rangle \sim t ^ \alpha,\quad \mbox{ for } \quad t<t_s,
\label{scalinglaw}
\end{equation}
where the scaling exponent changes from $\alpha\approx 1/4$ at larger quenching depths to $\alpha\approx 1/2$ at low undercooling rates. This dependence on the quenching temperature is also shown in  Fig.~\ref{scaling}. The late-stage evolution is characterized by a cross-over to a grain-growth stagnation regime where the steady-state mean grain area increases as the quenching depth is decreased. We observe that the cross-over time $t_s$ depends non-trivially on the system size $L$ and the quenching temperature, such that its value diverges as $L\rightarrow\infty$ and $T\rightarrow T_m$. In the inset of Fig.~\ref{scaling} we show the average grain area for stagnated states in a system of size $2048 \times 2048$ at different temperatures. We observe that the average grain area reaches the system size for $a_2\approx 0.05$ far below the melting temperature which in our case is equivalent to $a_2=0.25$. 

By tracking the angular evolution of the individual grains and their change in areas, we have considered in detail the correlation between the rotation rate and growth rate of the grains. In Fig.~\ref{graindist}, we show the grain trajectories in the space of their misorientation and area size. We notice that at the early stages in the coarsening process, the small grains tend to follow a random meandering in the misorientation space, i.e. small grain rotate much more than bigger grains. However, in the later coarsening stage, the grain boundary network is between large grains with selected misorientations. Also, we notice that some of the big grains suddenly disappear, which is an indication of coalescence where one grain rotates until his lattice orientation aligns with that of one of its neighbours. In the inset of Fig.~\ref{graindist}, we observe that for low temperatures the grain stagnation is concomitant with a rapid decrease in grain rotation. 

%
\begin{figure}
\includegraphics[width=.48\textwidth]{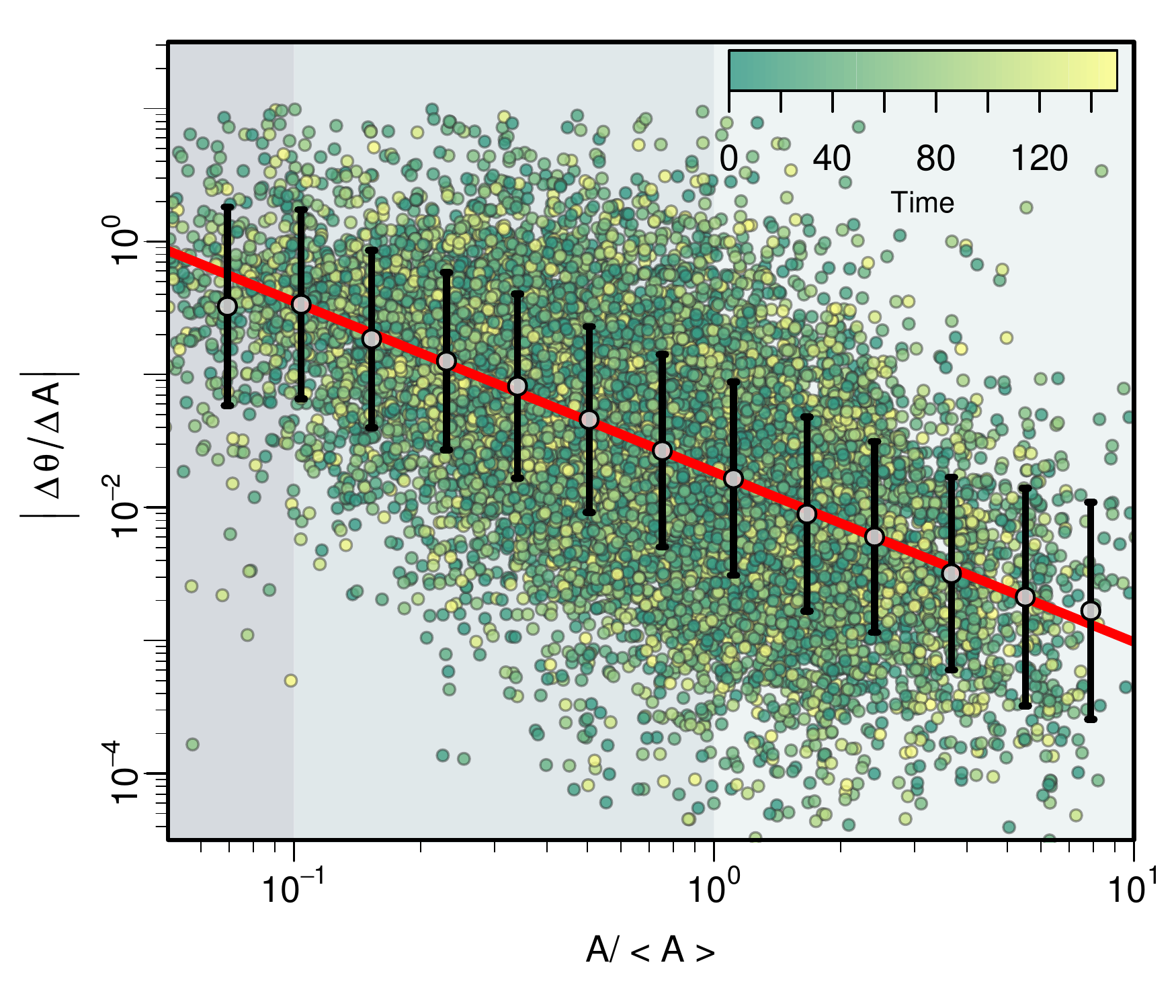}%
\caption{Grain rotation over area change $|\Delta \theta / \Delta A|$ in a fixed time step $\Delta t=1500$ is shown as function of the area divided by the mean area (at a given time step) $A/\langle A\rangle$. We observe a systematic decrease in the rotation per area as the grain area is reduced. The colors of the individual points, indicated by the color legend, represent the time at which a date point was observed. We note that there seems to be no clear difference in this plot between the early and late stage dynamics. The local average value of $|\delta \theta / \delta A|$ is shown together with the sample variation (standard deviation) marked by errorbars in black and the line on top is a best fit with a power law. The best fit yields an exponent $\beta=-1.25\pm 0.06$.}\label{dtheta}
\end{figure}

In general, the amount of grain rotation per change in area goes up for small grains, see Fig.~\ref{dtheta}. A best fit suggests, for relatively large areas, a scaling relation between the grain misorientation change and area change on the form
\begin{equation}
 \left|\frac{\Delta \theta}{\Delta A}\right| \sim A^{-\beta},
\label{thetascaling}
\end{equation}
where the scaling exponent is estimated to be $\beta=-1.25\pm .06$ for larger quenching depths ($a_2=-0.05$), and for simulations at low quenching depths the exponent $\beta$ slightly increases to a value $\beta=-1.1\pm .05$ at ($a_2=0.15$). If the grain rotation is solely due to the coupling to the normal motion, i.e. $rd\theta/dt = \Gamma v_n$, the conservation of the number of dislocations along the grain boundary implies that $r(t)\theta(t)=const$, or equivalently $\theta(t)\sim A^{-1/2}(t)$~\cite{cahn2004unified}. Consequently, it follows that $\beta = 3/2$. We ascribe the difference between the measured and predicted value of $\beta$ to the variation in misorientation that a grain has in a polyscrystalline matrix with its neighbors. At low quenching, the crystals are softer and grain boundary network more~\lq greased\rq~ allowing for sliding, dislocation reactions or even premeltings. When we lower the quenching depth (raising $a2>0.15$) most of the energy is dissipated in grain growth rather than grain rotation (see Fig.~\ref{graindist}, inset). Equivalently, grain rotation eventually stops sufficiently close to the melting temperature.

In summary, we have studied the anomalous coarsening and grain growth stagnation in polycrystalline films using the phase field crystal model. We find that the coarsening law is characterized by a power-law increase of the mean grain area with an exponent that depends on the quenching temperature and system size. This suggests that the sub-diffusive coarsening law is non-universal, although the probability distribution of grain size has shown to be robust against temperature variations. Moreover, we observe that the late stage coarsening is accompanied by a sudden decrease in grain rotation and crosses over to a stagnation regime with an average grain size that depends on temperature. 

{\it Acknowledgement} -- This study was supported through a grant "Earth Patterns" by the Villum Foundation and by Physics of Geological Processes.

\bibliographystyle{natbib}
\bibliography{stag}

\end{document}